\documentclass[12pt]{article}

\usepackage{amssymb}
\def\be{\begin{equation}}
\def\ee{\end{equation}}
\def\bqa{\begin{eqnarray}}
\def\eqa{\end{eqnarray}}
\def\roughly#1{\mathrel{\raise.3ex
\hbox{$#1$\kern-.75em\lower1ex\hbox{$\sim$}}}}

\begin{document}




\begin{center}
{\Large\bf Comment on\\ ``The New ${\mathbf F_L}$ Measurement from HERA and the
Dipole Model''}
\end{center}
\vspace {0.5 cm}
\begin{center}
{\bf Dieter Schildknecht\footnote[1]{Email: schild@physik.uni-bielefeld.de}} \\[
2.5mm]
Fakult\"{a}t f\"{u}r Physik, Universit\"{a}t Bielefeld \\[1.2mm] 
D-33501 Bielefeld, Germany \\[1.2mm]
and \\[1.2mm]
Max-Planck Institute f\"ur Physik (Werner-Heisenberg-Institut),\\[1.2mm]
F\"ohringer Ring 6, D-80805, M\"unchen, Germany
\end{center}

\vspace{1 cm}




\begin{abstract}
The upper bound on the ratio of the proton structure functions
$F_L/F_2$ tested in the recent paper ``The New $F_L$ Measurement
from HERA and the Dipole Model'', contrary to what is said therein, does
not provide a model-independent ``rigorous'' experimental test of the
color-dipole picture. The validity of the theoretical upper bound
depends on an ad hoc assumption on the dipole cross section. -- The
analysis in the paper ``The New $F_L$ Measurement
from HERA and the Dipole Model'' can be reinterpreted as an additional 
confirmation
of the absolute model-independent prediction from the color-dipole
picture of $F_L = 0.27 F_2$.
\end{abstract}







In a recent paper by Ewerz et al. \cite{Ewerz}, it is shown
that the experimental data \cite{Aaron} from HERA on the 
longitudinal-to-transverse ratio of the photoabsorption cross
sections in electron-proton deep inelastic scattering (DIS),
\be
R(W^2,Q^2) = \frac{\sigma_{\gamma^*_L p} (W^2,Q^2)}{\sigma_{\gamma^*_T p}
(W^2,Q^2)},
\label{1}
\ee
at large $Q^2$, or equivalently, the experimental data on the ratio
of the proton structure functions, $F_L(x,Q^2)/F_2(x,Q^2)$, reach values
close to, or even slightly above, the upper bound on $R(W^2,Q^2)$ previously
derived \cite{Nacht} within the color-dipole picture (CDP), compare
refs. \cite{Nikolaev} to \cite{MKDS}, of DIS at low
$x \cong Q^2/W^2 \ll 0.1$. It was noted that a violation of the ``rigorous''
upper bound on $R(W^2, Q^2)$ would falsify the validity of the CDP of DIS.

It was not explicitly stressed in ref. \cite{Ewerz}, however, that the
derivation \cite{Nacht} of the upper bound on $R(W^2,Q^2)$,
\be
R(W^2,Q^2) \le 0.37248 \cong 0.37,
\label{2}
\ee
or, equivalently, of the ratio of the proton structure functions,
\be
\frac{F_L (x,Q^2)}{F_2 (x,Q^2)} \le 0.27139 \cong 0.27,
\label{3}
\ee
relies on the {\it ad hoc} assumption \cite{Nacht} of the
dipole cross section being independent of the variable $0 \le z
\le 1$ that specifies the distributions of the momenta between
the quark and the antiquark in the $q \bar q$ color-dipole state
the photon fluctuates into. Without adopting this assumption, the proof
of the upper bound on $R (W^2,Q^2)$ breaks down.

The well-known example of a specific ansatz \cite{DIFF2000, KU-SCHI}
for the dipole cross section, to be referred to below, shows that a
$z(1-z)$-dependent dipole cross section {\it can} indeed lead to a
violation of the upper bound (\ref{2}) on $R(W^2,Q^2)$. A violation of
the ``rigorous'' upper bound by experimental data, accordingly, would only
rule out the use of the restrictive {\it ad hoc} ansatz of a 
$z(1-z)$-independent
dipole cross section within the CDP, rather than the CDP itself.

In what follows, we elaborate on the above statements.

We shall end by noting that the interesting detailed analysis of the
experimental data in ref. \cite{Ewerz}, demonstrating an approximate
saturation of the bound (\ref{3}), -- in contrast to the conclusions in
ref. \cite{Ewerz} --, confirms the validity of the CDP. As seen in
fig. 2 and fig. 4 in ref. \cite{Ewerz}, within errors, the experimental
data on $F_L/F_2$ provide additional confirmation of the 
model-independent absolute predictions in refs. \cite{MKDS, KU-SCHI}.

The photoabsorption cross section on protons at a given 
center-of-mass energy, $W$, for longitudinally and transversely polarized
photons, $\gamma^*_L$ and $\gamma^*_T$, of virtuality $Q^2$, in the CDP
in standard notation, takes the form, e.g. ref. \cite{KU-SCHI},
\be
\sigma_{\gamma^*_{L,T}p} (W^2, Q^2) = \int dz \int d^2 \vec r_\bot
\vert \psi_{L,T} (\vec r_\bot, z (1 - z), Q^2) \vert^2~~\sigma_{(q\bar q)p}
(\vec r_\bot, z (1 - z), W^2). 
\label{4}
\ee
According to (\ref{4}), the photon ``fluctuates'' into a $q \bar q$
state characterized by the transverse size $\vec r_\bot$ and the
configuration variable $z(1-z)$ with $0 \le z \le 1$, the $\gamma^*
\to q \bar q$ fluctuation probability being given by $\vert 
\psi_{L,T} (\vec r_\bot, z (1-z), Q^2) \vert^2$. The subsequent scattering
of the $q \bar q$ dipole state on the proton is described by the dipole
cross section $\sigma_{(q \bar q)p} (\vec r_\bot, z(1-z), W^2)$ in
(\ref{4}).

Under the {\it ad hoc} assumption that the dipole cross section in
(\ref{4}) be independent of the configuration variable $z(1-z)$,
\be
\sigma_{(q \bar q)p} (\vec r_\bot, z(1-z),W^2) = \sigma_{(q \bar q)p}
(\vec r_\bot, W^2),
\label{5}
\ee
the integration over $dz$ in (\ref{4}) can be carried out, and (\ref{4}) 
becomes
\be
\sigma_{\gamma^*_{L,T}p} (W^2,Q^2) = \int d^2 \vec r_\bot \omega_{L,T}
(\vec r_\bot, Q^2) \sigma_{(q \bar q)p} (\vec r_\bot, W^2),
\label{6}
\ee
where
\be
\omega_{L,T} (\vec r_\bot, Q^2) = \int dz \vert \psi_{L,T} (\vec r_\bot,
z(1-z), Q^2) \vert^2.
\label{7}
\ee
Rewriting the longitudinal photoabsorption cross section in (\ref{6}) in
terms of the transverse one via 
\be
\sigma_{\gamma^*_L p} (W^2,Q^2) = \int d^2 \vec r_\bot \omega_T 
(\vec r_\bot, Q^2) \frac{\omega_L (\vec r_\bot, Q^2)}{\omega_T 
(\vec r_\bot, Q^2)} \sigma_{(q \bar q)p} (\vec r_\bot, W^2),
\label{8}
\ee
and replacing the ratio of $\omega_L (\vec r_\bot, Q^2)/\omega_T 
(\vec r_\bot, Q^2)$ on the right-hand side in (\ref{8}) by its maximum
with respect to $0 \le \vec r_\bot \le \infty$ (and with respect to
the contributing quark flavors), one arrives \cite{Nacht} at an upper bound on
the ratio $R (W^2, Q^2)$ in (\ref{1}),
\be
R(W^2,Q^2) \le \max_{\vec r_\bot}
\frac{\omega_L (\vec r_\bot, Q^2)}
{\omega_T(\vec r_\bot, Q^2)}.
\label{9}
\ee
The well-known expression for the $\gamma^* \to q \bar q$ transition
amplitude can be used to evaluate the right-hand side in (\ref{9})
numerically. One finds \cite{Nacht} the above-mentioned numerical bound
(\ref{2}) on the ratio $R(W^2,Q^2)$ given by $R(W^2,Q^2) \le  0.37$.

It is clear that the proof of the bounds (\ref{2}) and (\ref{3}) 
crucially depends on the assumed independence (\ref{5}) of the dipole
cross section with respect to the variable $z (1-z)$. 

The specific
example of an ansatz for the 
dipole cross section, to which we turn
next, explicitly demonstrates that the bounds (\ref{2}) and (\ref{3}) 
{\it can} be violated, as soon as one removes the assumption (\ref{5})
of a $z (1-z)$-independent dipole cross section.

Consider the dipole cross section \cite{DIFF2000, KU-SCHI}
\bqa
\sigma_{(q \bar q)p} (\vec r_\bot, z (1-z), W^2)& = &
\sigma_{(q \bar q)p} (r^\prime_\bot, W^2)  \label{10}\\
& = &\sigma^{(\infty)} (W^2)
\left( 1 - J_0 \left( r_\bot \sqrt{z (1-z)} \Lambda_{sat} (W^2) \right)
\right).\nonumber
\eqa
In (\ref{10}), the variable $r^\prime_\bot$ stands for $r^\prime_\bot =
r_\bot \sqrt{z(1-z)}$, and $J_0 \left(r^\prime_\bot \Lambda_{sat} (W^2) 
\right)$ denotes the Bessel function with index 0. For definiteness, we 
note \cite{DIFF2000} that realistically the cross section $\sigma^{(\infty)} (W^2)$ has
to be of hadronic size, and approximately constant, while the
``saturation scale'' $\Lambda^2_{sat} (W^2)$ rises as a small power of
$W^2$.

The photoabsorption cross section (\ref{4}), restricting ourselves to
massless quarks, explicitly reads \cite{Nikolaev, KU-SCHI}
\bqa
& & \sigma_{\gamma^*_{L,T}p} (W^2, Q^2) = \frac{3\alpha}{2 \pi^2} \sum_q
Q^2_q Q^2 \cdot \label{11} \\
& & \hspace*{-0.5cm} \cdot \left\{ \matrix{ 
4 \int d^2 \vec r_\bot \int dz z^2 (1-z)^2  K^2_0 (r_\bot \sqrt{z(1-z)} Q)
\sigma_{(q \bar q)p} (r_\bot, z(1-z),W^2)\cr
\hspace{-6cm}    \int d^2 \vec r_\bot \int dz (1-2z (1-z)) z (1-z)}\right.
\nonumber \\
& &\cdot \hspace{4cm}  K^2_1 (r_\bot
\sqrt{z (1-z)}Q) \sigma_{(q \bar q)p} (r_\bot, z(1-z), W^2),
\nonumber
\eqa
where the upper and the lower line on the right-hand side refer to
longitudinally and transversely polarized photons, respectively, and
$K_0 (r_\bot \sqrt{z(1-z)} Q$) and $K_1 (r_\bot \sqrt{z(1-z)}Q)$ stand
for modified Bessel functions. The sum over the squares of the 
charges of the (actively contributing) quarks is denoted $\sum_q Q^2_q$,
and $\alpha$ is the electromagnetic fine-structure constant. With an
$r^\prime_\bot$-dependent dipole cross section, compare (\ref{10}),
upon introducing the variable $r^\prime_\bot = r_\bot \sqrt{z(1-z)}$ in
(\ref{11}), the $z(1-z)$ dependence factorizes. The integration over
$dz$ can be carried out to obtain \cite{KU-SCHI}
\be
\sigma_{\gamma^*_{L,T}p} (W^2, Q^2) = \frac{\alpha}{\pi} \sum_q Q^2_q Q^2
\int dr^{\prime 2}_\bot K^2_{0,1} (r^\prime_\bot Q) 
\sigma_{(q \bar q) p} (r^\prime_\bot, W^2).
\label{12}
\ee
In view of a later discussion, we note that only the first equality
in (\ref{10}) was used in the transition from (\ref{11}) to (\ref{12}).

As a consequence of the strong fall-off of the modified Bessel
functions for large values of their argument,
\be
K^2_{0,1} (r^\prime_\bot Q) \sim \frac{\pi}{2 r^\prime_\bot Q}
e^{-2 r^\prime_\bot Q}~~,~~~~~(r^\prime_\bot Q \gg 1)
\label{13}
\ee
the integral over $dr^{\prime 2}_\bot$ in (\ref{12}), to very good
approximation, is determined by the region of $r^{\prime 2}_\bot$ 
restricted by $r^{\prime 2}_\bot \ll 1/Q^2$. For large values of 
$Q^2$, it is sufficient to evaluate the photoabsorption cross section
(\ref{12}) upon substituting a suitable approximation of the dipole
cross section (\ref{10}) valid in the restricted domain of 
$r^{\prime 2}_\bot \ll 1/Q^2$.

For $r^{\prime 2}_\bot \ll 1/\Lambda^2_{sat} (W^2)$, the specific dipole
cross section of the second equality in (\ref{10}) can be approximated by
\be
\sigma_{(q \bar q)p} (r^\prime_\bot, W^2) = \sigma^{(\infty)} (W^2) 
\frac{1}{4} r^{\prime 2}_\bot \Lambda^2_{sat} (W^2) + \cdots ~~.
\label{14}
\ee
Chosing $Q^2$ sufficiently large, such that $1/Q^2 \ll 1/\Lambda^2_{sat}
(W^2)$, the range of $r^{\prime2}_\bot \ll 1/Q^2$ determining the
photoabsorption cross section (\ref{12}) lies within the range of
validity, $r^{\prime 2}_\bot \ll 1/\Lambda^2_{sat} (W^2)$, of the
approximation (\ref{14}) of the dipole cross section (\ref{10}).
Substituting (\ref{14}) into (\ref{12}), one finds that the
longitudinal-to-transverse ratio $R(W^2,Q^2)$ for the specific dipole 
cross section
(\ref{10}), at sufficiently large $Q^2$, is given by
\be
R(W^2,Q^2) = \frac{\int dr^\prime_\bot r^{\prime 3}_\bot K^2_0
(r^\prime_\bot Q)}{\int dr^\prime_\bot r^{\prime 3}_\bot K^2_1
(r^\prime_\bot Q)} = \frac{1}{2}.
\label{15}
\ee
The result $R(W^2,Q^2) = 0.5$ violates the bound (\ref{2}) that was
obtained under the restrictive assumption (\ref{5}) of a
$z(1-z)$-independent dipole cross section.

A possible violation of the bound (\ref{2}) by experimental data
does not rule out the validity of the CDP. A violation of the bound
(\ref{2}) only rules out the restrictive {\it ad hoc} assumption (\ref{5})
of a $z(1-z)$-independent dipole cross section.

In ref. \cite{Nacht}, the crucial assumption (\ref{5}) was motivated
by the requirement of a factorization of longitudinal ($z$-dependence)
and transverse ($r_\bot$ dependence) degrees of freedom in high-energy
reactions. As seen
in (\ref{11}), even upon adopting the assumption (\ref{5}) on the
dipole cross section, there is no factorization in the expression
for the photoabsorption cross section, since, even under the
assumption (\ref{5}), there is a remaining dependence on the product
$r_\bot \sqrt{z(1-z)}$ on the right-hand side in (\ref{11}).

Instead of the transverse variable $\vec r_\bot$, without loss of
generality, equivalently, one may formulate the CDP, compare (\ref{4})
and (\ref{11}), in terms of the transverse variable $\vec r^{~\prime}_\bot
= \vec r_\bot \sqrt{z(1-z)}$ by carrying out the substitution 
$\vec r_\bot = \vec r^{~\prime}_\bot/\sqrt{z(1-z)}$ in (\ref{11}).
Requiring factorization of the $z(1-z)$ dependence in this 
$r^\prime_\bot$-representation of the CDP amounts to introducing the 
assumption
\be
\sigma_{(q \bar q)p} \left( \frac{r^\prime_\bot}{\sqrt{z(1-z)}}, 
z(1-z), W^2 \right) = \sigma_{(q \bar q)p} (r^\prime_\bot, W^2)
\label{15a}
\ee
in replacement of (\ref{5}). Since (\ref{15a}) coincides with the first
equality in (\ref{10}), upon substitution of (\ref{15a}) into (\ref{11})
and integration over $dz$, we obtain (\ref{12}). Applying 
an argument analogous to the one that led from (\ref{6}) to (\ref{9}), 
from (\ref{12}), one
finds the bound
\be
R(W^2,Q^2) \le \max_{r^\prime_\bot} \frac{K^2_0 (r^\prime_\bot Q)}
{K^2_1 (r^\prime_\bot Q)} = 1
\label{15b}
\ee
The weaker bound (\ref{15b}), based on (\ref{15a}), compared with the stronger
bound (\ref{2}), based on (\ref{5}), explicitly demonstrates that a
violation of the stronger bound (\ref{2}) by experimental data is 
consistent with the CDP, even upon supplementing the CDP with a 
factorization principle, as advocated for in \cite{Nacht}. The factorization
principle does not require the $z(1-z)$ independence (\ref{5}) that
leads to the bound (\ref{2}). The assumption (\ref{5}) is an {\it ad hoc}
assumption. A violation of the bound (\ref{2}) by experimental data
would neither violate the CDP nor the factorization principle.

Rewriting (\ref{11}) in terms of the transverse-size variable
$r^\prime_\bot = r_\bot \sqrt{z(1-z)}$, without introducing the
assumption (\ref{15a}), one obtains \cite{KU-SCHI}
\be
\sigma_{\gamma^*_{L,T}p} (W^2, Q^2) = \frac{\alpha}{\pi} \sum_q Q^2_q Q^2
\int dr^{\prime 2}_\bot K^2_{0,1} (r^\prime_\bot Q) 
\sigma_{(q \bar q)^{J=1}_{L,T} p} (r^\prime_\bot, W^2).
\label{16}
\ee
No specific assumption on the dipole cross section is needed to arrive
at the factorized form of the photoabsorption cross section in (\ref{16}).
Equation (\ref{16}) represents
\cite{KU-SCHI} the photoabsorption cross section in terms of the
$\gamma^*_L \to (q \bar q)^{J=1}_L$ and $\gamma^*_T \to (q \bar q)^{J=1}_T$
transition probabilities for fixed transverse size $r^\prime_\bot$, multiplied
by the respective cross sections for longitudinally and transversely
polarized dipole states, $\sigma_{(q \bar q)^{J=1}_L p} (r^\prime_\bot, W^2)$ 
and 
$\sigma_{(q \bar q)^{J=1}_T p} (r^\prime_\bot, W^2)$,
on the proton. 

The color-gauge-invariant interaction of the $q \bar q$ dipole state with
the gluon field of the nucleon implies ``color transparency'' \cite{Nikolaev},
the vanishing of the dipole cross section with vanishing dipole size, 
$r^2_\bot \to 0$ in (\ref{4}), and $r^{\prime 2}_\bot \to 0$ in (\ref{16}).
In this limit, the ratio, $\rho_W$, of the dipole cross sections 
$\sigma_{(q \bar q)^{J=1}_T p} (r^\prime_\bot, W^2)$ and 
$\sigma_{(q \bar q)^{J=1}_L p} (r^\prime_\bot, W^2)$ in (\ref{16}) is
independent of the dipole size. At most, it depends on the energy, $W$,
\be
\frac{\sigma_{(q \bar q)^{J=1}_T p} (r^{\prime 2}_\bot, W^2)}{\sigma_{(q \bar q
)^{J=1}_L p} (r^{\prime 2}_\bot, W^2)} = \rho_W~ ,~~~~ ({\rm for}~ 
r^{\prime 2}_\bot \to 0).
\label{17}
\ee
Evaluation of (\ref{16}), by taking into account color transparency, and 
introducing the ratio (\ref{17}), then implies a longitudinal-to-transverse
ratio $R (W^2, Q^2)$ given by \cite{MKDS, KU-SCHI}
\be
R(W^2, Q^2) = \frac{1}{2 \rho_W}.
\label{18}
\ee
The gauge-invariant interaction of the $q \bar q$ dipole with the gluon
field of the nucleon, at sufficiently large $Q^2$, implies the 
longitudinal-to-transverse ratio (\ref{18}), with so far undetermined 
proportionality factor $\rho_W$ from (\ref{17}).

Dipole states, $(q \bar q)^{J=1}_T$, 
originating from a transversely polarized photon, $\gamma^*_T$,
differ in their internal quark (antiquark) transverse momentum distribution
from dipole states,$(q \bar q)^{J=1}_L$, originating from a
longitudinally polarized photon,
$\gamma^*_L$.
The relatively small transverse quark momentum in a transversely polarized
dipole state, $(q \bar q)^{J=1}_T$, relative to a longitudinally polarized
one, $(q \bar q)^{J=1}_L$, via the
uncertainty principle, implies a relatively large transverse size of the 
$(q \bar q)^{J=1}_T$ state compared with the $(q \bar q)^{J=1}_L$ state.
Quantitatively, one finds a definite value for the size enhancement. The
proportionality factor in (\ref{17}) is given by  
the $W$-independent constant of magnitude \cite{MKDS, KU-SCHI, Ringberg11}
\be
\rho_W = \rho = \frac{4}{3}.
\label{19}
\ee

With (\ref{19}), the ratio $R(W^2,Q^2)$ in (\ref{18}) is uniquely predicted
by
\be
R(W^2,Q^2) = \frac{1}{2 \rho} = \frac{3}{2 \cdot 4} = 0.375.
\label{20}
\ee
In terms of the proton structure functions, (\ref{20}) becomes
\be
F_L (x \simeq Q^2/W^2, Q^2) = 0.27 F_2 (x \simeq Q^2/W^2, Q^2).
\label{21}
\ee

The result (\ref{21}) is a unique, model-independent
 consequence of the color-gauge-invariant
interaction of a $q \bar q$ dipole state with the gluon field in the
nucleon as formulated within the CDP.

The approximate numerical agreement of the equalities (\ref{20}) and
(\ref{21}) with the upper bounds in (\ref{2}) and (\ref{3}) is purely
accidental.

We briefly return to the {\it ad hoc} ansatz for the dipole cross section in
(\ref{10}). By comparison of (\ref{16}) with (\ref{12}), we conclude that
the ansatz (\ref{10}) contains the assumption of helicity independence
for $(q \bar q)^{J=1}$ dipole-proton scattering,
\be
\sigma_{(q \bar q)^{J=1}_L p} (r^\prime_\bot, W^2) = 
\sigma_{(q \bar q)^{J=1}_T p} (r^\prime_\bot, W^2) = \sigma_{(q \bar q)p}
(r^\prime_\bot, W^2).
\label{22}
\ee
Assumption (\ref{22}) replaces the transverse-size-enhancement factor
$\rho = 4/3$ in (\ref{19}) by unity, $\rho = 1$, thus ignoring the
different 
internal structure of $(q \bar q)^{J=1}_L$ and $(q \bar q)^{J=1}_T$
states.

In ref. \cite{KU-SCHI}, by comparison with the experimental data, it was
concluded that the prediction (\ref{21}), within errors, agrees with the
experimental results from HERA.

The conclusion in ref. \cite{KU-SCHI} is confirmed by the recent results
in ref. \cite{Ewerz}. As a consequence of the approximate numerical 
coincidence of the equality (\ref{21}) with the upper bound (\ref{3}), the
results in fig. 2 and in fig. 4 in ref. \cite{Ewerz} may be reinterpreted
as an experimental test of the equality (\ref{21}). Within errors, the
comparison with the results from HERA \cite{Aaron}, shown in the aforementioned
figs. 2 and 4 in ref. \cite{Ewerz}, confirms the conclusion \cite{KU-SCHI}
on the consistency with experiment of the model-independent prediction 
(\ref{21}) of the CDP.

\bigskip

{\Large\bf Acknowledgement}

The results of the present comment are largely based on the quoted
published work in
collaboration with Kuroda-san.








\end{document}